\documentclass{article}

\usepackage[round]{natbib}

\usepackage{multiequations}
\usepackage{graphicx}
\usepackage{xcolor}
\usepackage{amsfonts}
\usepackage{amsmath}
\usepackage{amssymb}
\usepackage{empheq} 
\usepackage[hidelinks]{hyperref}
\usepackage{enumerate}  
\usepackage[loadonly]{enumitem}  
\usepackage{bbm}

\usepackage{geometry}

\newcommand*{\be}{\begin{equation}}
\newcommand*{\ee}{\end{equation}}
\newcommand*{\bea}{\begin{eqnarray}}
\newcommand*{\eea}{\end{eqnarray}}
\newcommand*{\bal}{\begin{align}}
\newcommand*{\eal}{\end{align}}
\newcommand*{\bme}{\begin{multiequations}}
\newcommand*{\eme}{\end{multiequations}}

\newcommand{\defi}{\stackrel{\triangle}{=}}
\newcommand{\dv}{\bnabla\! \bcdot\!}

\newcommand{\id}{\mbs{\mathbb{I}}_d}
\providecommand\bnabla{\boldsymbol{\nabla}}
\providecommand\bcdot{\boldsymbol{\cdot}}

\newcommand\Ros{\mbox{\textit{R\scriptsize o}}}  
\newcommand\Bu{\mbox{\textit{B\scriptsize u}}}  
\newcommand\Fr{\mbox{\textit{F\scriptsize r}}}  
\newcommand\Eu{\mbox{\textit{E\scriptsize u}}}  

\renewcommand*{\Omega}{\varOmega}
\renewcommand*{\Sigma}{\varSigma}

\newsavebox{\astrutbox}
\sbox{\astrutbox}{\rule[-5pt]{0pt}{20pt}}

\def\squarebox#1{\hbox to #1{\hfill\vbox to #1{\vfill}}}

\newcommand{\Dt}{\mathbb D}

\newcommand{\w}{\boldsymbol{w}}

\newcommand{\B}{\boldsymbol{B}}
\newcommand{\bsigma}{\boldsymbol{\sigma}}

\newcommand{\xx}{\boldsymbol{x}}
\newcommand{\kk}{\boldsymbol{k}}
\newcommand{\XX}{\boldsymbol{X}}

\newcommand{\nab}{\boldsymbol{\nabla}}

\newcommand{\transp}{^{\scriptscriptstyle T}}

\newcommand{\Exp}{\mathbb{E}}

\newcommand{\dif}{{\mathrm{d}}}
\newcommand{\mbs}[1]{\ensuremath{\boldsymbol{#1}}}

\newcommand{\dBt}{\dif  \boldsymbol{B}_t}

\begin{document}

%

\title{Geophysical flows under location uncertainty, Part III\\
SQG and frontal dynamics under strong turbulence conditions}


\author{V. Resseguier${\dag\ddag}$$^{\ast}$\thanks{$^\ast$Corresponding author. Email: valentin.resseguier@inria.fr
\vspace{6pt}},
 E. M\'emin${\dag}$
and B. Chapron${\ddag}$
\\\vspace{6pt}  ${\dag}
$Inria, Fluminance group, Campus universitaire de Beaulieu, Rennes Cedex 35042, France\\ ${\ddag}
$Ifremer, LOPS, Pointe du Diable, Plouzan\'e 29280, France
}

\maketitle

\begin{abstract}

Models under location uncertainty are derived assuming that a component of the velocity is uncorrelated in time. The material derivative is accordingly modified to include an advection correction, inhomogeneous and anisotropic diffusion terms and a multiplicative noise contribution. This change can be consitently applied to all fluid dynamics evolution laws. This paper continues to explore benefits of this framework and consequences of specific scaling assumptions. Starting from a Boussinesq model under location uncertainty, a model is developed to describe a mesoscale flow subject to a strong underlying submesoscale activity. As obtained, the geostrophic balance is modified and the Quasi-Geostrophic (QG) assumptions remarkably lead to a zero Potential Vorticity (PV). The ensuing Surface Quasi-Geostrophic (SQG) model provides a simple diagnosis of warm frontolysis and cold frontogenesis.


\end{abstract}

\section{Introduction}

Quasi-Geostrophic (QG) models are standard models to study mesoscale barotropic and baroclinic dynamics. Assuming uniform Potential Vorticity (PV) in the fluid interior, the Surface Quasi-Geostrophic (SQG) model helps describe the surface dynamics \citep{blumen1978uniform,Held95,Lapeyre06,constantin1994formation,constantin1999front,constantin2012new}.  Despite its simplicity, the SQG relation provides a good diagnosis to relate mesoscale surface buoyancy fields to surface and interior velocity fields. Nevertheless, QG and SQG paradigms assume strong rotation and strong stratification ($\Fr\sim \Ros \ll1$) and thus neglect the submesoscale ageostrophic dynamics. In particular, the QG velocity is horizontal and solenoidal. This structure prevents the emergence and development of realistic submesoscale features such as frontogenesis, restratification, and asymmetry between cyclones and anticyclones  \citep{lapeyre2006oceanic,Klein08}. In contrast, the QG$^{+1}$ \citep{muraki1999next} and SQG$^{+1}$ \citep{hakim2002new} models capture such phenomenon with a (one degree) higher order power series expansions in the Rossby number. This comes with an additional complexity. In particular, the SQG$^{+1}$ model involves a nonlinear PV. Semi-Geostrophic (SG) \citep{eliassen1949quasi,hoskins1975geostrophic} and Surface Semi-Geostrophic (SSG) models \citep{hoskins1976baroclinic,hoskins1979baroclinic,badin2013surface,ragone2016study} also offer simple alternatives to the QG framework. Within a weaker stratification context ($\Fr^2\sim \Ros \ll 1$), ageostrophic terms emerges to  better represent fronts and filaments than QG dynamics.
The SSG model is formally similar to SQG as it is in the same way associated with a zero PV. Yet, SSG involves a space remapping (from geostrophic coordinates to physical coordinates)  together with a nonlinear term in the PV that is often neglected \citep{ragone2016study}. These terms -- both of order $1$ in Rossby -- bring relevant horizontal velocity divergence as in SQG$^{+1}$ model. Nevertheless, these terms require a more involved numerical inversion.

In this paper, we derive a linear SQG model enabling to cope with frontal dynamics  without explicitly resolving higher Rossby order. PV is not arbitrarily set to zero, it rigorously results from a strong submesoscale activity.
Such a derivation is a direct consequence of the dynamics under location uncertainty \citep{Memin14,resseguier2016geo1,resseguier2016geo2}, for which the velocity is decomposed between a large-scale resolved component and a time-uncorrelated unresolved component. 
Derived models then rigorously handle sub-grid tensors. In particular, they link together small-scale velocity statistics, turbulent diffusion, small-scale induced velocity and backscattering effects.

After briefly recalling the main features of models under location uncertainty (section 2), a modified SQG model is derived (section 3). Finally, the ensuing diagnostic relation is tested on realistic very-high resolution model outputs (section 4).

\section{Models under location uncertainty}
\label{Models under location uncertainty}

Hereafter, we briefly outline the main ideas for the derivation of these stochastic models (for a more complete description, see \cite{resseguier2016geo1}). This relies on a decomposition of the flow velocity in terms of a large-scale component, $\w$, and a random field uncorrelated in time, $\bsigma \dot{\B}$:
\bea
\label{decompo flow}
\frac{\dif \XX}{\dif t} &=& \w + \bsigma \dot{\B}.
\eea
The latter represents the small-scale velocity component. This solenoidal, possibly anisotropic and non-homogeneous random field corresponds to the aliasing effect of the unresolved velocity component. To parametrize its spatial correlations, an infinite-dimensional linear operator, $\bsigma$, is applied to a space-time white noise, $\dot{\B}$. The decomposition \eqref{decompo flow} leads to a stochastic representation of the Reynolds transport theorem (RTT) and of the material derivative, $D_t$ (derivative along the flow \eqref{decompo flow}). In most cases, this derivative coincides with the stochastic transport operator, $\Dt_t$, defined for every field, $\Theta$, as follow:
\bea
{ \Dt}_t \Theta \ 
&\defi&
\underbrace{
\dif_t \Theta
}_{
\substack{
 \defi \  \Theta(\xx,t+\dif t) - \Theta(\xx,t) \\
\text{Time increment}
}
}
+ 
\underbrace{
\left ({\w}^\star\dif t + \bsigma \dif\B_t \right)\bcdot \nab \Theta
}_{\text{Advection}}
 - 
\underbrace{
\nab\bcdot \left ( \frac{1}{2} \mbs a  \nab \Theta \right )
}_{\text{Diffusion}}
\dif t ,
\label{Mder}
\eea 
where the time increment term $\dif_t \Theta$ stands instead of the partial time derivative as $\Theta$ is non differentiable. The diffusion coefficient matrix, $\mbs a$, is solely defined by the one-point one-time covariance of the unresolved displacement per unit of time:
\bea 
\mbs a = 
\bsigma \bsigma \transp =
\frac{
\Exp \left \{ \bsigma \dBt \left( \bsigma \dBt  \right)\transp  \right \} 
}{\dif t}
,
\label{balance}
\eea
and the modified drift is given by
\bea
\w^\star = \w   - \frac{1}{2} ( \nab\bcdot \mbs a)\transp
.
\eea
The stochastic RTT and material derivative involve a diffusive subgrid term, a multiplicative noise and a modified advection drift induced by the small-scale inhomogeneity.
%
This material derivative has a remarkable conservative property. Indeed, for any field, $\Theta$, randomly transported, {\em i.e.}
\bea
\Theta (\XX(t + \Delta t),t + \Delta t) 
&=&
 \Theta (\XX(t),t),
 \eea
  \cite{resseguier2016geo1} showed that the energy of each realization is conserved:
\bea
\frac{\dif}{\dif t} \int_\Omega \Theta^2 =0.
\eea

The RTT enables us to express the conservation law of mechanics (linear momentum, energy, mass) with a partially known velocity. Deterministic and random subgrid parametrizations for various geophysical flow dynamics can then directly be obtained. Stochastic Navier-Stokes and Boussinesq models can be derived as discussed by \cite{Memin14} and \cite{resseguier2016geo1}. The latter model involves random transports of buoyancy and velocity, together with incompressibility constraints.

\section{Mesoscale flows under strong uncertainty}
\label{subsection Simplified stochastic oceanic models}


From the Boussinesq model, the QG assumptions state a strong rotation and a strong stratification. This is of particular interest to study flows at mesoscale, where both kinetic and buoyant dynamics are important. More specifically, we focus on horizontal length scales, $L$, such as:
\begin{eqnarray}
\label{QG scalings}
\frac 1 {\Bu} 
= \left(  \frac {\Fr}{\Ros} \right)^2
= \left ( \frac L {L_d} \right )^2 
\sim 1
\text{ and }
 \frac 1 \Ros = \frac{L f_0} U \gg 1,
\end{eqnarray}
where $U$ is the horizontal velocity scale, $L_d \defi \frac{Nh}{f}$ is the Rossby deformation radius, $N$ is the stratification (Brunt-V\"ais\"al\"a frequency) and $h$ is the characteristic vertical length scale. In the following, both differential operators Del, $\nab$, and Laplacian, $\Delta$, represent $2$D operators. Moreover, $\bsigma_{H \bullet} $ stands for the horizontal component of $\bsigma$, $\mbs a_H$ for $\bsigma_{H \bullet} \bsigma_{H \bullet}  \transp$ and $A_u$ for its scaling.




\subsection{Specific scaling assumptions}

Similarly to \cite{resseguier2016geo2}, scalings within the QG framework \eqref{QG scalings} can authorize the  set up of a non-dimensional stochastic Boussinesq model amenable to further simplifications.

\subsubsection{Quadratic variation scaling}
\label{Scaling}

Models under location uncertainty involve subgrid terms which have also to be scaled. A new dimentionless number, $\Upsilon$, quantifying the ratio of horizontal advection and horizontal turbulent diffusion is therefore introduced:
\bea
\Upsilon
\defi  \frac{U/L}{A_u/L^2 }
= \frac{U^2}{A_u / T}.
\eea
We can also relate it to the ratio of Mean Kinetic Energy (MKE), $U^2$, to the Turbulent Kinetic Energy (TKE), $A_u/T_{\sigma}$, where $T_\sigma$ is the small-scale correlation time. This reads:
\bea
\Upsilon
= \frac{1}{\epsilon}\frac{MKE}{TKE} 
,
\eea
where $\epsilon=T_\sigma/T$ is the ratio of the small-scale to the large-scale correlation times.
This parameter, $\epsilon$, is central in homogenization and averaging methods \citep{Majda99,givon2004extracting,Gottwald13}. The number $\Upsilon/ \Ros$ measures the ratio batween rotation and horizontal diffusion. For a parameter $\Upsilon$ close or larger than unity, the geostrophic balance still holds \citep{resseguier2016geo2}, whereas for $\Upsilon\sim \Ros $, this balance is modified.

The parameter $\Upsilon$ depends through $A_u$ on the flow and on the resolution scale. In order to specifiy the scaling and the resulting associated model, knowledge of the characteristic horizontal eddy diffusivity or eddy viscosity is needed. Tuning experiences of usual subgrid parametrizations may provide such information, and \cite{boccaletti2007mixed} give some examples of canonical values.

If absence of characteristic values, absolute diffusivity or similar mixing diagnoses could be measured \citep{keating2011diagnosing} as a proxy of the variance tensor. Small values of $\Upsilon$ are generally relevant for the ocean where the TKE is often one order of magnitude larger than the MKE \citep{wyrtki1976eddy,richardson1983eddy,stammer1997global,Vallis}.
Note that here the TKE may encompass all the unresolved dynamics down to the Kolmogorov scale.

\subsubsection{Vertical unresolved velocity}
\label{vertical velocity Scaling}
To scale the vertical unresolved velocity, we consider

\begin{eqnarray}
\label{scaling_vertical_sigma_dBt}
\frac { (\bsigma \dif \B_t)_{z} }{\| (\bsigma \dif \B_t)_{H}\| } \sim \frac{\Ros}{\Bu} D,
\end{eqnarray}
where $D=\frac h L$ is the aspect ratio and the subscript $H$ indicates horizontal coordinates. 
The $\omega$-equation \citep{giordani2006advanced} justifies such a scaling. For any velocity $\mbs{\mathfrak{u}} = (\mbs{\mathfrak{u}}_H,\mathfrak{w})\transp$, which scales as $(\mathfrak{U},\mathfrak{U},\mathfrak{W})\transp$, this equation reads
\begin{equation}
f_0^2 \partial_z^2 \mathfrak{w}
+ N^2 \Delta \mathfrak{w}
=\nab \bcdot \mbs Q
\approx
- \nab \bcdot \left (  \nab \mbs{\mathfrak{u}}_H \transp  \nab \mathfrak{b} \right)
\approx
- f_0\nab \bcdot  \left (  \nab  \mbs{\mathfrak{u}}_H  \transp \partial_z \mbs{\mathfrak{u}}_H^{\bot} \right),
\end{equation}
where  $\mathfrak{b}$ stands for the buoyancy variable and $\mbs Q$ for the so-called $\mbs Q$-vector. In its non-dimentional version, the $\omega$-equation reads:
\begin{equation}
\frac{\mathfrak{W}}{\mathfrak{U}}
\left (
\partial_z^2 \mathfrak{w}
+ \Bu \Delta \mathfrak{w}
\right)
\approx 
D \Ros
\nab \bcdot \mbs Q.
\end{equation}
The Burger number is small at planetary scales where the rotation dominates ($\frac{\mathfrak{W}}{\mathfrak{U}} \sim D \Ros 
$) and is large at submesoscales where the stratification dominates ($\frac{\mathfrak{W}}{\mathfrak{U}} \sim D \Ros/\Bu
$).
For the small-scale velocity $\bsigma \dot{\mbs B}$, the latter is thus more relevant.

Relations between the isopicnical tilt and mixing give another justification of the scaling \eqref{scaling_vertical_sigma_dBt}. Based on baroclinic instabilities theory, anisotropy specifications of eddy diffusivity sometimes rely on this tilt \citep{Vallis}. Moreover, several other authors suggest that the eddy activity and the associated mixing mainly occur along isentropic surfaces \citep{Gent90,Pierrehumbert93}.

For QG dynamics, the Burger is of order one and the scaling in $D \Ros$ and in $D\Ros/\Bu$ coincides. In particular, they encode a mainly horizontal unresolved velocity:
\begin{eqnarray}
\label{scaling_vertical_sigma_dBt_QGmoderate_uncertainty}
\frac { (\bsigma \dif \B_t)_{z} }{\| (\bsigma \dif \B_t)_{H}\| } \sim \frac{\Ros}{\Bu} D \ll D.
\end{eqnarray}
This is consistent with the assumption of a large stratification, {\em i.e.} flat isopycnicals, if we admit that the activity of eddies preferentially apprears along the isentropic surfaces.
As a consequence, the terms $(\bsigma \dif \B_t)_z\partial_z$ scale as $\frac{\Ros}{\Bu} (\bsigma \dif \B_t)_H \bcdot \nab$. In the QG approximation, the scaling of the diffusion and effective advection terms including $\bsigma_{z \bullet}$ are one to two orders smaller (in power of $\Ros/\Bu $) than terms involving $\bsigma_{H\bullet}$. For any function $\xi$, the vertical diffusion $\partial_z( \frac{\bsigma_{z \bullet} \bsigma_{z \bullet} \transp} 2  \partial_z \xi)$ is one order smaller than the horizontal-vertical diffusion term $ \nab \bcdot \left( \frac{\mbs \bsigma_{H \bullet} \bsigma_{z \bullet} \transp} 2 \partial_z \xi \right)$ and two orders smaller than the horizontal diffusion term $ \nab \bcdot \left( \frac{\bsigma_{H \bullet} \bsigma_{H \bullet} \transp} 2 \nab \xi \right)$.

\subsubsection{Beta effect}
\label{beta effect Scaling}
The beta effect is weak at mid-latitude mesoscales. Yet, at the first order, it influences the absolute vorticity. So, we choose the same scaling as \cite{Vallis}:
\begin{eqnarray}
\beta y \sim \nab^{\bot} \bcdot \mbs u \sim \frac U L = \Ros f_0.
\end{eqnarray}


\subsection{Stratified Quasi-Geostrophic model under strong uncertainty}
\label{subsubsection Stratified Quasi-Geostrophic model under strong uncertainty}

Strong uncertainty condition corresponds to $\Upsilon$ having an order of magnitude close to the Rossby number. More specifically, we assume $\Ros \leqslant \Upsilon \ll 1$. In this situation, the random eddies have larger energy than the large-scale mean kinetic energy. 
Accordingly, the diffusion and drift terms are one order of magnitude larger than the advection terms.


In the case of strong ratio $\Upsilon$, the diffusion is very large and the system is not approximately in geostrophic balance anymore. The large-scale horizontal velocity becomes divergent, and decoupling the system is more tedious. For sake of simplicity, in the following we consider the case of homogeneous and horizontally isotropic turbulence. As a consequence, the variance tensor, $\mbs a$, is constant in space and diagonal:
\begin{equation}
\mbs a =
\begin{pmatrix}
a_H & 0 & 0 \\
0 & a_H & 0 \\
0 & 0 & a_z
\end{pmatrix}.
\end{equation}

\subsubsection{Modified geostrophic balance under strong uncertainty}

For horizontal homogeneous turbulence, the large-scale geostrophic balance is modified by the horizontal diffusion, whereas the unresolved velocity is in geostrophic balance:
\begin{subequations}
\label{eqGeoModifeq:tot}
\begin{empheq}
[left={}
\empheqlbrace]{align}
  & 
\mbs f \times \mbs u - \frac{a_H} 2 \Delta \mbs u 
= - \frac 1 {\rho_b} \nab p' ,
  \label{eqGeoModifeq:largeScale} \\
  & 
 \mbs f \times 
\bsigma_H \dBt 
 = - \frac 1 {\rho_b}\nab \dif_t p_{\sigma} 
,
  \label{eqGeoModifeq:smallScale}
\end{empheq}
\end{subequations}
where $\mbs u$ is the large-scale horizontal velocity, $p'$ the time-correlated component of the pressure, $\dot{p}_\sigma=\frac{\dif p_\sigma}{\dif t}$ the time-uncorrelated component, and $\rho_b$ is the mean density.
For a constant Coriolis frequency, the first equation can be solved in Fourier space. The Helmholtz decomposition of the velocity reads:
\begin{subequations}
\label{eqGeoModif:tot}
\begin{empheq}
[left={}
\empheqlbrace]{align}
  & 
  \mbs{u} =
 \nab^{\bot} \psi + \nab \tilde{\psi} ,
  \label{eqGeoModif:helmholtz} \\
  & 
  \hat{\psi} =
 \left( 1 +  \left \| \frac {\mbs k} {k_c} \right \|^4_2\right)^{-1}
  \frac  {\hat{p}'} 
  {\rho_b f},
  \label{eqGeoModif:sol}\\
  &
  \tilde{\psi} =
  \frac {1} {k_c^2} \Delta \psi,
  \label{eqGeoModif:div}
\end{empheq}
\end{subequations}
where $k_c = \sqrt{\frac{2f_0}{a_H}}$ and the hat accent indicates a horizontal Fourier transform. This solution is derived in Appendix \ref{appendix Modified geostrophic balance} using geometric power series of matrices. The obtained formula is valid for any right hand side in equation \eqref{eqGeoModifeq:largeScale}. For instance, additional forcing such as an Ekman stress could be taken into account. In equation (\ref{eqGeoModif:tot}), the solenoidal component of the velocity, $\nab^{\bot} \psi$, corresponds to the usual geostrophic velocity multiplied by a low-pass filter \eqref{eqGeoModif:sol}. The irrotational (ageostrophic) component of the velocity, $\nab \tilde{ \psi}$, dilates the anticyclones (maximum of pressure and negative vorticity) and shrinks the cyclones (minimum of pressure and positive vorticity) at small scales. Indeed, according to equation (\ref{eqGeoModif:div}), the divergence of the velocity corresponds to the vorticity Laplacian divided by $k_c^2$. This realistic ageostrophic behavior is also present in the simplified surface oceanic models SQG$^{+1}$ \citep{hakim2002new} and Surface Semi-Geostrophic (SSG) \citep{badin2013surface,ragone2016study}.
  In the proposed stochastic model, the divergent component and the low-pass filter of the system (\ref{eqGeoModif:tot}) are parameterized by the spatial cutoff frequency $k_c$, which moves toward larger scales when the diffusion coefficient $a_H$ increases. If both the vorticity and the divergence can be measured at large scales, the previous relation should enable to estimate the cutoff frequency $k_c$ by fitting terms of equation (\ref{eqGeoModif:div}). Then, the horizontal diffusive coefficient, $a_H$, or the variance of the horizontal small-scale velocity (at the time scale $\Delta t$), $a_H/ \Delta t$, can be deduced.

\subsubsection{Modified SQG relation under strong uncertainty}

To derive a QG model, we use the other equations of the stochastic Boussinesq model at the $0$-order. After some algebra (see Appendix \ref{appendix QG model under strong uncertainty}), we obtain directly a zero PV in the fluid interior:
\begin{eqnarray}
PV = 
\left( \Delta
+  
\left( 1 + \frac{\Delta^2}{k_c^4} \right )
\partial_z
\left(
\left ( \frac{f_0}{N} \right)^2 
\partial_z
\right)
\right)
\psi
=  0,
\end{eqnarray}
where $k_c = \sqrt{\frac{2 f_0 }{a_H}}$.  The assumptions used here correspond to the same used for a classical QG model \citep{Vallis}, except that the dissipation, due to the noise, is strong. It is a striking result. Instead of finding a model in the form of a classical QG model, developments, through a strong uncertainty, directly leads to the description of surface dynamics, a SQG model. It means that the subgrid dissipation prevents the development of the interior dynamics. Without this dynamics, no baroclinic instabilities can grow \citep{Lapeyre06}. If the stratification is vertically invariant, this static linear equation can be solved by imposing a vanishing condition in the deep ocean ($z \to - \infty$) and a specified boundary value at a given depth ($z=\eta$). The horizontal Fourier transform of the solution then reads:
\begin{equation}
\hat \psi (\mbs k,z)
=
\hat \psi (\mbs k,\eta) \ 
\exp \left(
\frac{ N \| \mbs k \|_2}
{f_0
\sqrt{ 1 +  \left \| \frac {\mbs k} {k_c} \right \|^4_2 }
}
\ 
(z-\eta)
\right).
\end{equation}
At $z=\eta$, the modified SQG relation is:
\begin{equation}
\label{SQG relation under strong uncertainty}
\hat b(\mbs k , \eta)
=
N \| \mbs k \|_2
\sqrt{ 1 +  \left \| \frac {\mbs k} {k_c} \right \|^4_2 }
\ 
\hat \psi (\mbs k , \eta)
,
\end{equation}
where $b$ stands for the buoyancy. In the following, we will refer to \eqref{SQG relation under strong uncertainty} as the SQG relation under Strong Uncertainty ($SQG_{SU}$).
For low wave number or moderate uncertainty ($\| \mbs k/k_c \|^2_2\sim \Ros/\Upsilon\ll 1 $), we retrieve the standard SQG relation. 
The expression of the stream function as a function of the buoyancy is expressed as the convolution with a Green function, $G_{SQG}= \frac 1 {2 \pi N} \|\xx \|^{-1}$, and the velocity decays rapidly as the inverse of the square distance to the point vortex center.  On the other hand, for very high wave numbers or very large uncertainty ($\| \mbs k/k_c \|^2_2\sim \Ros/\Upsilon \gg 1 $), the velocity tends very quickly to zero. For strong uncertainty or small scales ($\| \mbs k/k_c \|^2_2\sim \Ros/\Upsilon \sim 1 $), 
\begin{eqnarray}
\hat b
=\frac{\sqrt{2}N}{k_c}
\left( \| \mbs k \|_2^2 
+ \underset{\| \mbs k \| \to k_c }{O}
\left( \left\|\frac{\mbs k }{k_c} \right \|_2 -1 \right)^2
\right) \hat{\psi}.
\end{eqnarray}
Accordingly, we may see the $SQG_{SU}$ relation as an intermediary between two relevant models in geophysics: the SQG dynamics where the tracer (the buoyancy) is proportional to $\| \mbs k \|_2 \hat{\psi}$ and a two-dimensional flow dynamics where the tracer (the vorticity) is proportional to $\| \mbs k \|_2^2 \hat{\psi}$.  In the latter case, the streamfunction can be expressed as the convolution of the buoyancy with the Green function, $G_{2D}= \frac {k_c} {2\sqrt{2} \pi N} \ln \|\xx \|$, and the velocity decays slowly as the inverse of the distance to the point vortex center. Nevertheless, contrary to the two-dimensional flow and the SQG models, the 2D velocity $\mbs u$ is divergent (see equation \eqref{eqGeoModif:div}). The total horizontal velocity can be computed from the buoyancy, through the Helmholtz decomposition \eqref{eqGeoModif:helmholtz}, the modified SQG relation \eqref{SQG relation under strong uncertainty} and the equation \eqref{eqGeoModif:div}. As derived, the vertical velocity is finite and given by the main balance of the buoyancy equation:
\begin{equation}
w 
= \frac {f_0}{N^2 } 
 \frac 1 {k_c^2} 
\Delta  b.
\label{vert-velocity}
\end{equation}
Note that this equation is not derived from a non-hydrostatic vertical momentum equation. 
Equation (\ref{vert-velocity}) is directly obtained from the thermodynamic equation. It expresses the fact that, under strong stratification and strong horizontal diffusion, the buoyancy anomalies are mainly created by vertical advection. This relation is similar to the result of \cite{garrett1981dynamical}, except the proportionality coefficient. Indeed, \cite{garrett1981dynamical} consider vertical diffusion and neglect the horizontal one. Invoking the thermal wind relation and the stratification structure, vertical variations are then associated with horizontal buoyancy variations. 
In the present development, the vertical velocity scales as $\frac{\Ros }{ \Upsilon \Bu} DU 
\sim \| \mbs k/k_c\|^2_2 \frac{DU}{\Bu} 
$. It is prominent at small scales and proportional to the variance tensor, such as the divergent component of the horizontal velocity.\\
 
\begin{figure}
	\centering
	\includegraphics[width=6in]{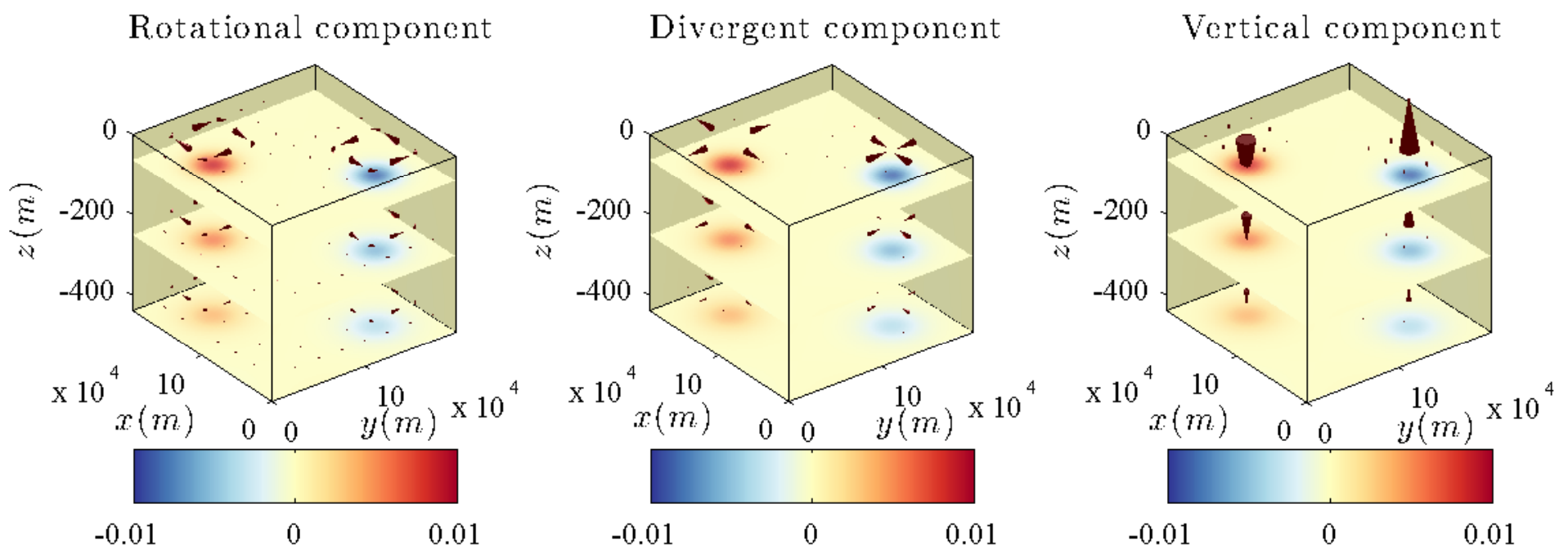}
	\caption{Value of the interior buoyancy created by a warm spot and a cold spot at the surface. The two components of the velocity are also shown. The upwelling and shrinking of the cyclone (cold spot) and the downwelling and spreading of the anticyclone (warm spot) are clearly visible.}
	\label{fig interior of SQG modif}
	\centering
	\includegraphics[width=6in]{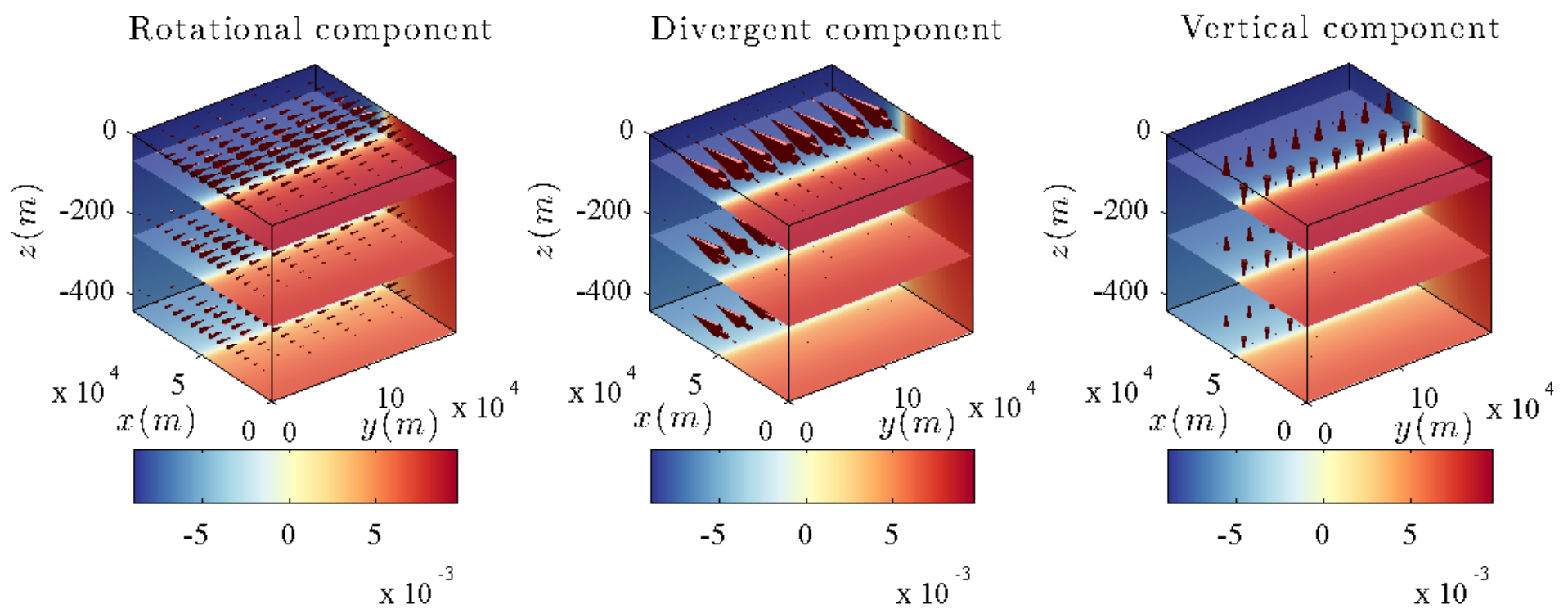}
	\caption{Value of the interior buoyancy created by a front at the surface. The two components of the velocity are also shown. The divergence effects will strengthen the front on the cold side (frontogenesis) and smooth the front on the warm side (frontolysis). The vertical velocity is here much weaker than in the case of isolated spots.}
	\label{fig interior of SQG modif front}
\end{figure}

Figures \ref{fig interior of SQG modif} and \ref{fig interior of SQG modif front} show the static link between the $3$D velocity and buoyancy for two isolated vortices and a front, respectively. 
As obtained, the solenoidal component is similar to the classic SQG velocity. In Figure \ref{fig interior of SQG modif}, the non-rotational component forces the anticyclone (warm spot) to spread, and the cyclone (cold spot) to shrink. Note that our study focuses on the ocean dynamics. For atmospheric applications, the vertical axis should be inverted and the sign of the temperature anomaly changed \citep{ragone2016study}. In Figure \ref{fig interior of SQG modif front}, the irrotational component is weak on the warm side of the front, but strongly strengthens the cold side. As modeled in SQG$^{+}$ and SSG, a frontolysis (resp. frontogenesis) develops on the warm (resp. cold) side of the front. In Figure \ref{fig interior of SQG modif}, a downwelling of warm water and a upwelling of cold water appear. As the vertical velocity comes from the thermodynamic equation and not from the vertical momentum equation, it is the cause of the buoyancy anomaly not its consequence. Whereas the irrotational horizontal component is stronger close to a front than within an eddy, the vertical velocity associated with a front is found much weaker than the one associated with an isolated eddy. 

\section{Diagnostic under strong uncertainty}
\label{SQG model under strong uncertainty}
As derived, under strong uncertainty, the eddy diffusion is substantial and modifies the geostrophic balance \eqref{eqGeoModif:tot}. The velocity becomes divergent and equation \eqref{eqGeoModif:div} offers a diagnostic of this divergence. This diagnostic states that the divergence should be proportional to the Laplacian of the vorticity:
\bea 
\delta  = \frac{1}{k_c^2} \Delta \zeta
\ \text{ where } \  
 \left \{
   \begin{array}{r c l}
      \delta &=& \dv \mbs u, \\
      \zeta   & = & \nab^{\bot} \bcdot \mbs u.
   \end{array}
   \right .
 \eea

To evaluate the relevance of this diagnostic, outputs of a realistic   $3$D  high-resolution oceanic simulation are used. During winter, the eddy activities are usually stronger, especially close to energetic currents. For this reason, the Gulf-Stream during winter season is a test-bed region for high-resolution simulation \citep{gula2015topographic}.

\begin{figure}
\begin{center} 
\includegraphics[width=7.85cm]{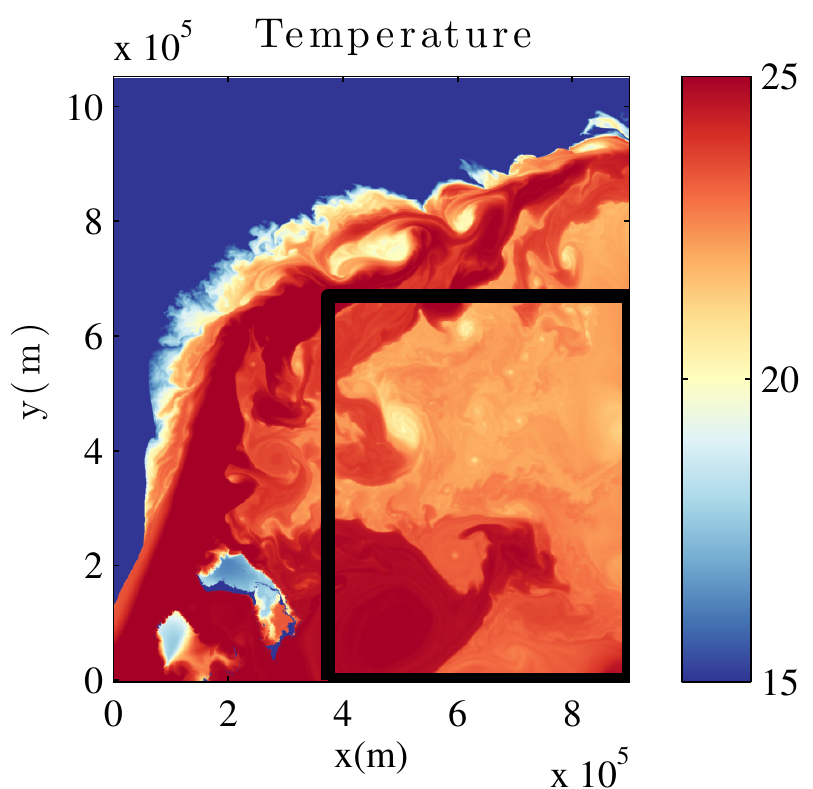}
\includegraphics[width=7.15cm]{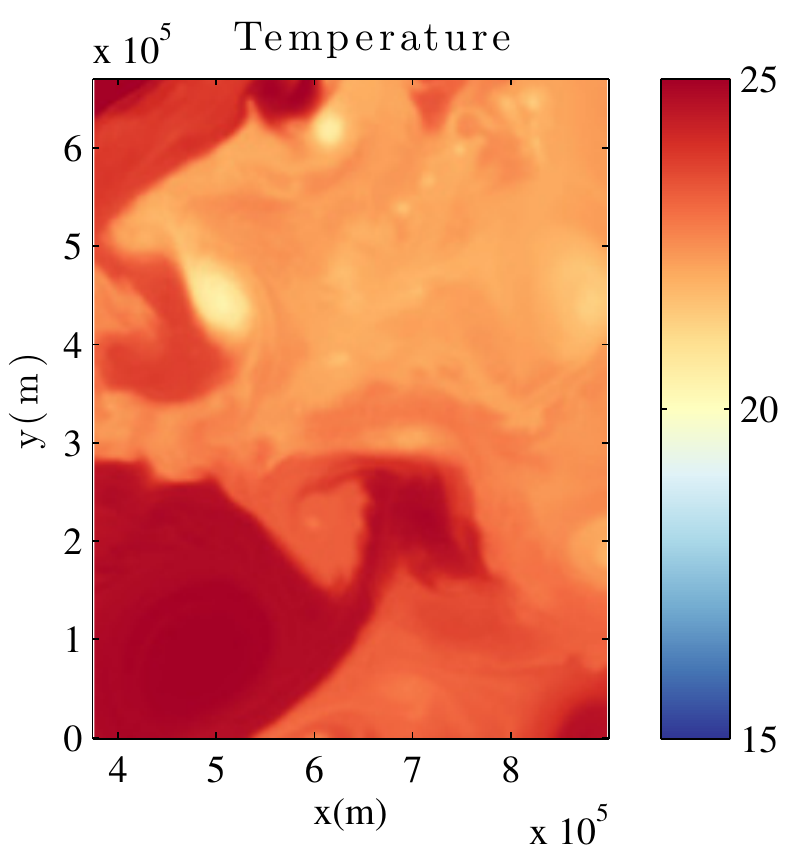}
\includegraphics[width=7.85cm]{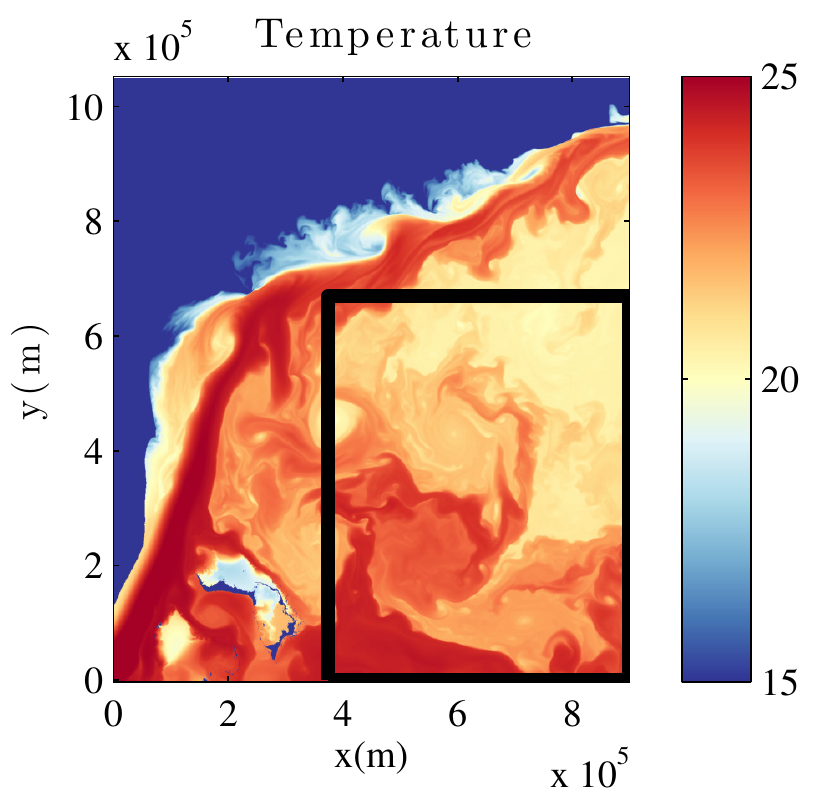}
\includegraphics[width=7.15cm]{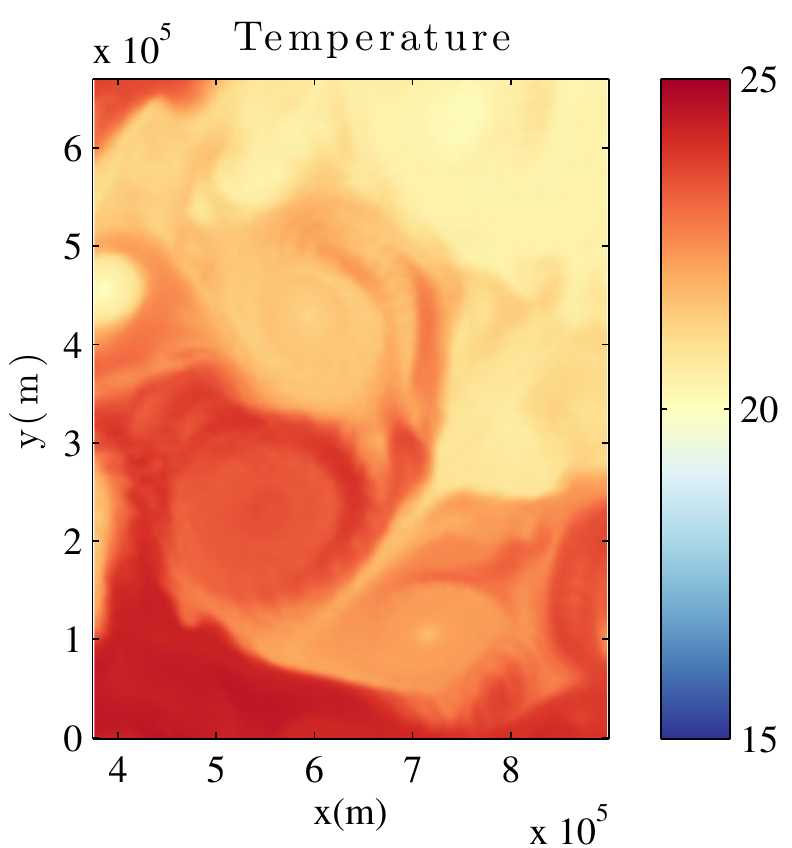}
\end{center}
  \caption{Temperature (in Celsius degree) for the first (top) and $58^{th}$ day (bottom) at high temporal and spatial resolution ($\Delta t =12$h and $\Delta x =750$m) (left) and after filtering ($\Delta t =1$ day and $\Delta x =3$km) (right). The black line on the top pictures highlight the region selected for the diagnostic.}
\label{plot_data_gula}
\end{figure}

Figure \ref{plot_data_gula} shows the temperature of the first and of the $58^{th}$ day. Simulations are three-dimensional and involve a fine spatial and temporal resolutions. 
Equation \eqref{eqGeoModif:div} is a surface mesoscale diagnostic valid far from the coasts. Consequently, the surface fields are filtered temporally and spatially. The final time step is one day and the final resulting spatial resolution is $3$ km. 
Figure \ref{plot_data_gula} displays the original surface field and the filtered cropped fields. 

\begin{figure}
\begin{center} 
\includegraphics[width=15cm]{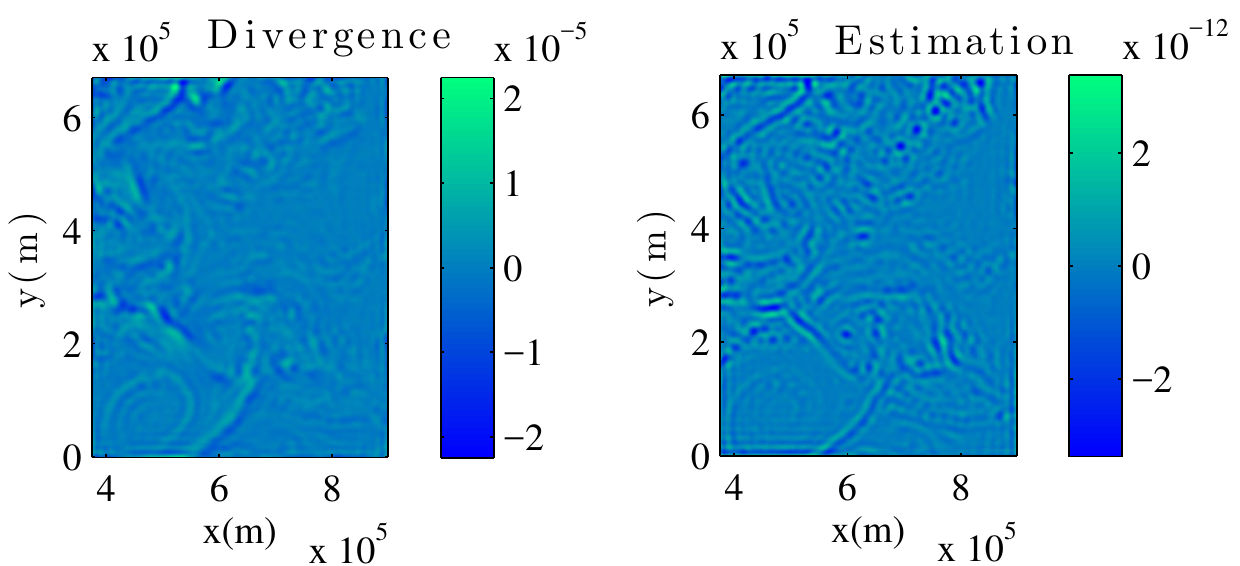}
\includegraphics[width=15cm]{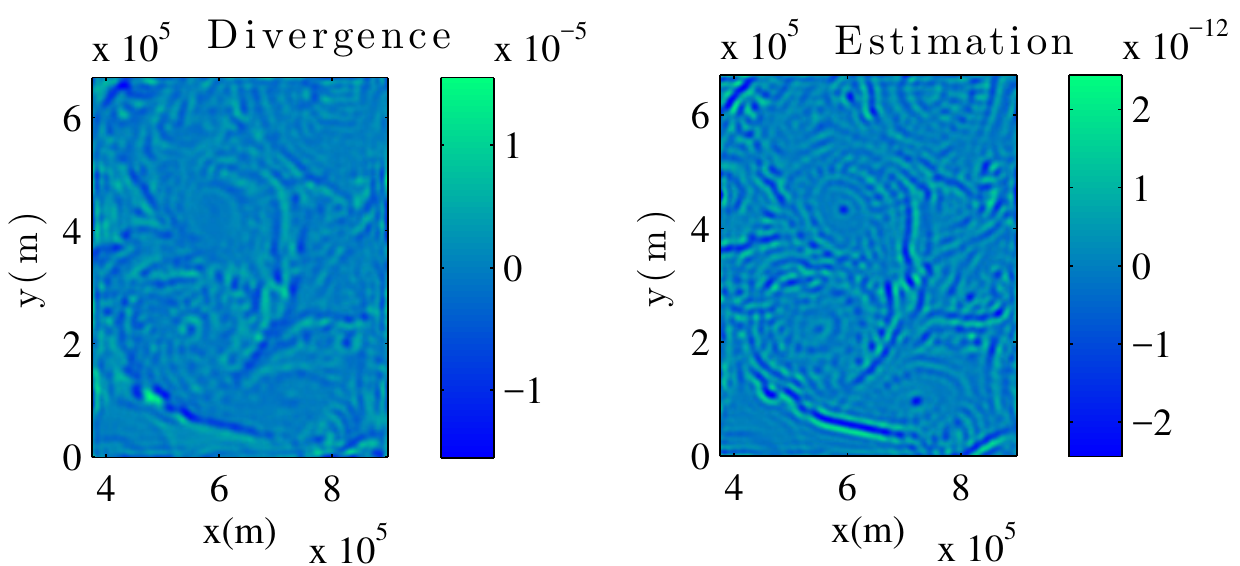}
\end{center}
\caption{Divergence ($s^{-1}$) and Laplacian of the vorticity ($m^{-2}.s^{-1}$) for the first and the $58^{th}$ day at a $30$-km resolution. According to our modified geostrophic balance under strong uncertainty, the latter is an estimation of the mesoscale divergence up to a multiplicative constant.
}
\label{spatial_diagno_div}
\end{figure}

Figure \ref{spatial_diagno_div} compares the reference divergence field to our estimate, the Laplacian of the vorticity. An overall agreement clearly emerges.
Nonetheless, the small scales of our estimate are more energetic than the small scales of the real divergence field. For this reason, the spatial fields are further filtered at a resolution of $30$ km. Except for some small spots, estimation and reference are similar. In particular, fronts -- associated with two length scales: one at sub-mesoscales and one at mesoscales -- are highlighted. 

\begin{figure}
\begin{center} 
\includegraphics[width=15cm]{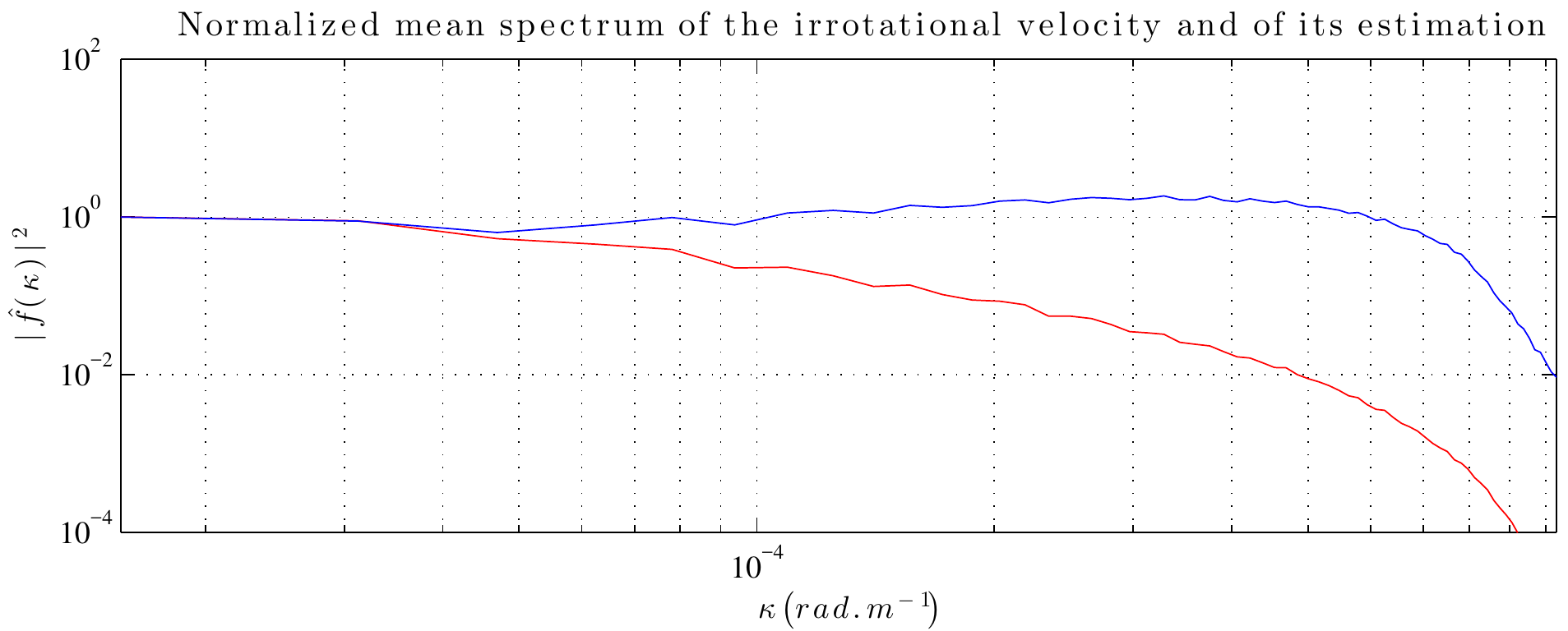}
\includegraphics[width=15cm]{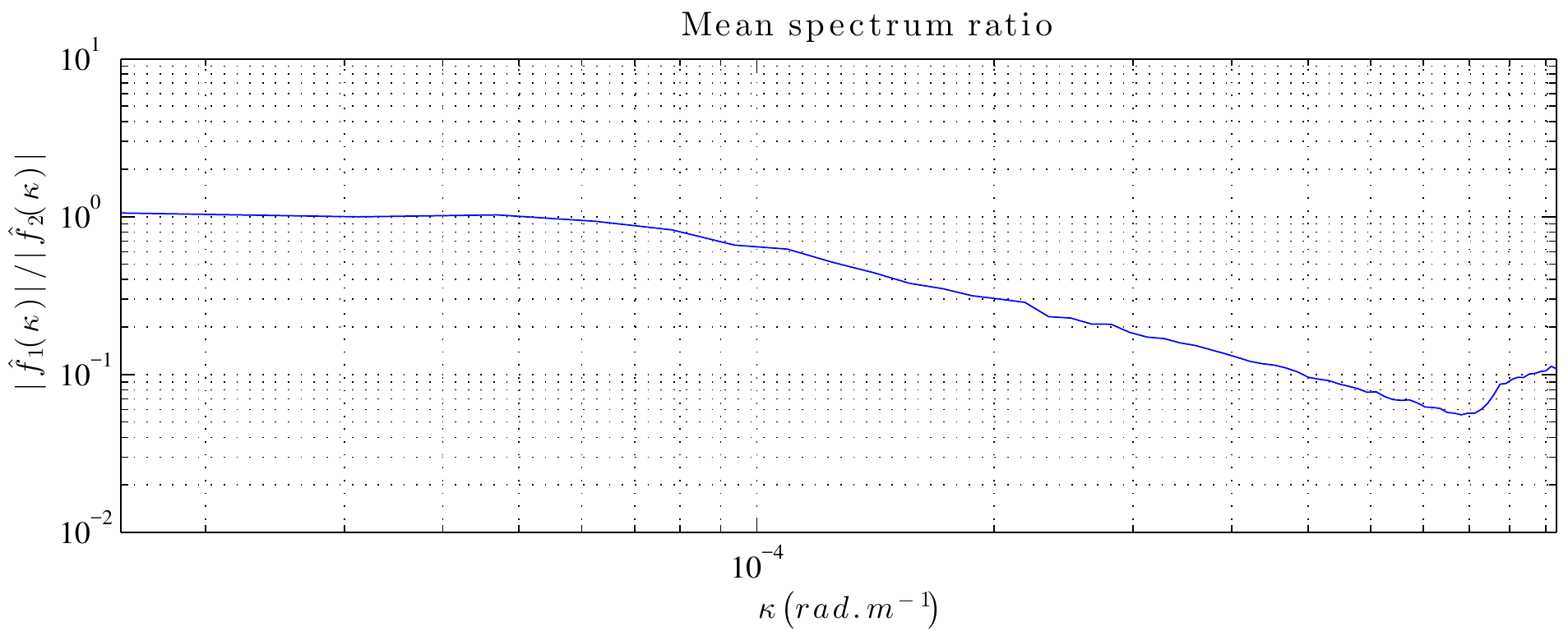}
\includegraphics[width=15cm]{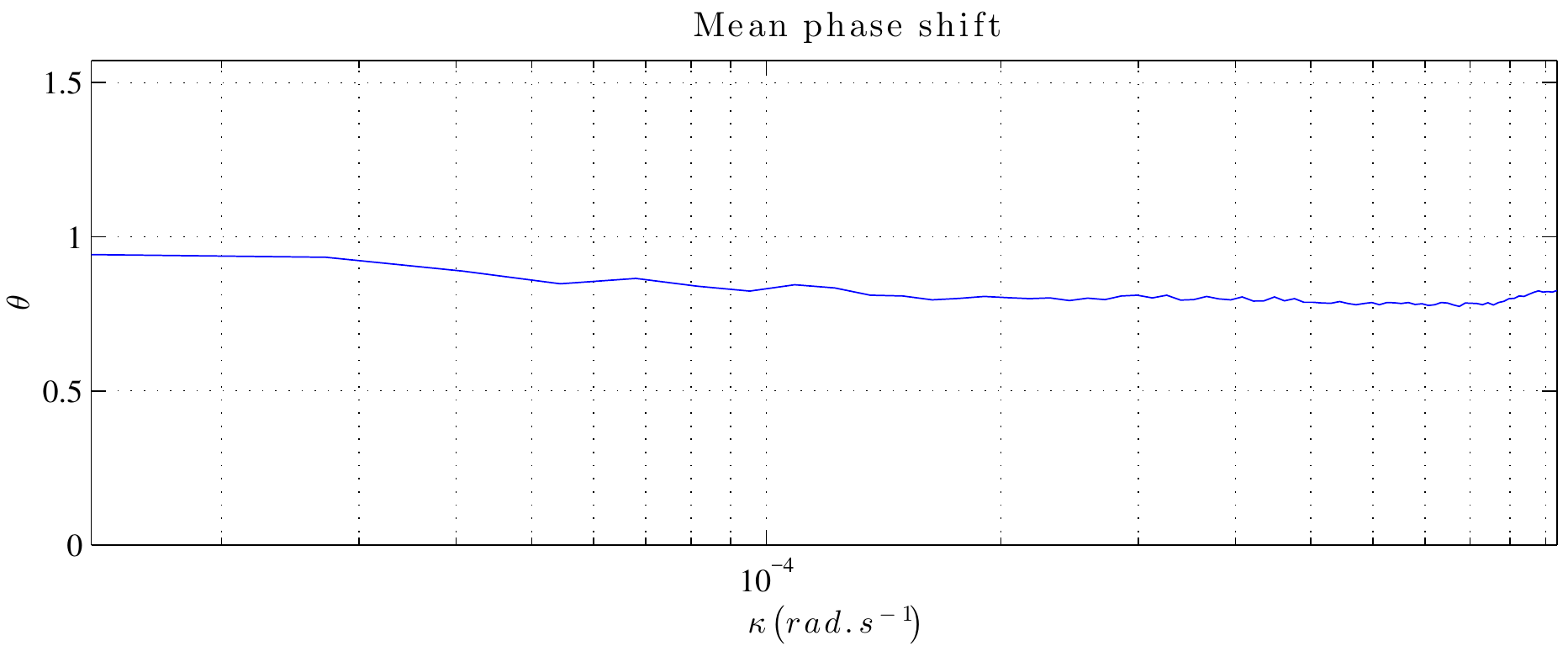}
\end{center}
\caption{From top to bottom: Normalized spectrum of the irrotational velocity component (red) and of our estimate of this component (blue), ratio of the Fourier transform modulus of the divergence to the one of our estimate and phase shift (rad) between the divergence and its estimate. Each of these spectral quantities is averaged on angular spatial frequencies and on the $58$ winter days.}
\label{spectrum_diagno_div}
\end{figure}

Figure \ref{spectrum_diagno_div} specifies the relevance and the limitations of the proposed diagnostic. The spectra of the two fields unveil a very good match at mesoscale range ($L>60$km {\em i.e.} $\kappa<10^{-4}$), whereas they differ at sub-mesoscales. This difference is certainly not surprising, the estimation being derived for large scale components. Note, the velocity divergent component is far from being zero in the mesoscale range. Compared to the solenoidal component, its spectrum is certainly much flatter and smaller in this range. Nevertheless, the mesoscale divergence is stronger than the sub-mesoscales divergence. The ratio of Fourier transform modulus further confirms the accuracy of our diagnostic at mesoscales and makes clear the difference at sub-mesoscales. The $-1$ slope may suggest that a fractional diffusion would be preferable to a Laplacian diffusion at those scales.

The complementary analysis is the coherence, which is a measure of the phase relationship between two fields. Specifically, the coherence is the Fourier modes correlation coefficient:
\bea
\Re \left (
\frac{
\widehat{\delta}(\kk)  \ 
\overline{\widehat{\Delta \zeta }(\kk)}
}{
\big |
\widehat{\delta}(\kk)  \ 
\overline{\widehat{\Delta \zeta }(\kk)}
\big |
}
\right)
,
 \eea
where $\Re$ denotes the real part. The coherence is the cosinus of the phase shift, $\theta$, between the two fields. Here, we directly show the phase shift averaged on angular spatial frequencies.

For our estimate the phase-shift is about $0.8 \approx \frac \pi 4 $. It means that a linear transformation of the large-scale vorticity can explain more than half of the divergence. As a comparison, the same analysis was done with the SQG relation, using temperature anomaly instead of buoyancy (not shown). The phase shift was similar. 

From Figure \ref{spatial_diagno_div}, one further get a rough estimation for the multiplicative constant of the proposed diagnostic: $k_c^2 \approx 10^{-7}$. It suggests a spatial cutoff $k_c^{-1} \approx 3$ km and a diffusion coefficient $a_H/2 \approx 1000 \ m^2.s^{-1}$. This value is canonical, according to \cite{boccaletti2007mixed}, which upholds the proposed approach. To confirm the validity of our strong uncertain assumption, it can be evaluated:
\bea 
\frac{\Ros}{\Upsilon}
\sim \left \| \frac{\mbs k }{k_c} \right \|^2_2
\sim  \frac{a_H }{2 f_0} \kappa^2
\sim 0.1, 
\eea 
with $\frac{ 2 \pi }{ \kappa } = 60$ km.

The unresolved energy can also be estimated. From a mesoscale point of  
view, motions induced with diurnal cycles can be approximated as  
delta-correlated processes. Hence, an estimation of the unresolved  
horizontal velocity amplitude shall follow from $\sqrt{a_H / \Delta t  
}\approx 10^{-1} m.s^{-1}$, with $\Delta t = 1$ day. Considering the  
present simulation, this is consistent with the sub-mesoscale velocity  
field.

\section{Conclusion}

To develop models under location uncertainty, the highly-oscillating unresolved velocity component is assumed to be uncorrelated in time. Consequently, the expression of the material derivative and hence most fluid dynamics models are modified, taking into account an inhomogeneous and anisotropic diffusion, an advection correction and a multiplicative noise. In this work, we simplify a Boussinesq model under location uncertainty assuming strong rotation, stratification, and subgrid turbulence. From this last assumption, the geostrophic balance is modified, and an horizontal divergent velocity explicitly appears. Furthermore, the QG approximation implies a zero PV. In other words, the strong uncertainty prevents interior dynamics at mesoscales. This provides a new derivation of the SQG model from the Boussinesq equations. The ensuing SQG model with divergent velocity is denoted $SQG_{SU}$. It exhibits physically relevant asymmetry between cold and warm areas, and suggests a diagnostic of the mesoscale divergence from the vorticity, as successfully tested on very high-resolution simulated data.
\\


A more complete model could encompass white noise components for temperature, salinity and density. At mesoscales, a thermal wind relation should relate these time uncorrelated components to the unresolved velocity. Therefore, these additional terms should provide the vertical structure of the unresolved velocity, without increasing the complexity of the parametrization.


Finally, besides solar forcing, the restratification is certainly a complicated process related to frontal dynamics. In the Mixed Layer (ML), the ML instabilities are often triggered by non-hydrostatic motions. They generate very-small-scale baroclinic instabilities and slumpings of the fronts \citep{boccaletti2007mixed}. For such phenomena, subgrid parameterizations are necessary. They must act to horizontally homogenize and restratify the ML. In such a context, the $SQG_{SU}$ model may constitute a simple solution or, at least a first step to develop  models under location uncertainty in this direction. To encode the weak stratification of the ML, stochastic Semi-Geostrophic (SG) and Surface Semi-Geostrophic (SSG) models could also be derived. According to our scaling of the vertical unresolved velocity \eqref{scaling_vertical_sigma_dBt}, a weaker stratification should then enhance the vertical mixing compared to the $SQG_{SU}$ model.


\section*{Acknowledgments}

The authors thank Aur\'elien Ponte, Jeroen Molemaker and Jonathan Gula for helpful discussions. We also acknowledge the support of the ESA DUE GlobCurrent project, the ``Laboratoires d'Excellence''  CominLabs, Lebesgue and Mer through the SEACS project.


\bibliographystyle{plainnat}

\bibliography{biblio}


\appendix


\section{Modified geostrophic balance}
\label{appendix Modified geostrophic balance}

Under strong horizontal homogeneous turbulence, the large-scale geostrophic balance is modified by the horizontal diffusion:
\begin{eqnarray}
\mbs f \times \mbs u - \frac{a_H} 2 \Delta_H \mbs u = \mbs \xi,
\end{eqnarray}
where $\mbs u$ is the resolved horizontal velocity and $\Delta_H \defi \partial_x^2 + \partial_y^2$ the horizontal Laplacian.
On the right-hand side, $\mbs \xi$ is the pressure gradient. 
Let us note that $\mbs f \times \mbs u = f \mbs J \mbs u$ with $
\mbs J 
=
\begin{pmatrix}
0 & -1\\
1 & 0
\end{pmatrix}
$ and that $\mbs J \transp = \mbs J^{-1} = - \mbs J$.
For a constant Coriolis frequency, the previous equation can be solved in the horizontal Fourier space :
\begin{eqnarray}
\hat{ \mbs u}
 = \left( f \mbs J + \frac{a_H} 2 \| \mbs k \|_2^2 \id \right )^{-1}
 \hat {\mbs \xi}
 = \left( \id - \left \| \frac{\mbs k}{k_c} \right \|_2^2 \mbs J \right )^{-1}
  \left( - \frac 1 f \hat {\mbs \xi}^{\bot} \right),
\end{eqnarray}
with $k_c =  \sqrt{ \frac{2 f}{a_H}}$. $ - \frac 1 f \mbs \xi^{\bot} = - \frac 1 f \mbs J \mbs \xi$ is the solution without diffusion. Expanding the right-hand side operator in Taylor series and using the properties $\mbs J^{2p}=(-1)^p \id$ and $\mbs J^{2p+1}=(-1)^p \mbs J$,
\begin{eqnarray}
 \left( \id - \left \| \frac{\mbs k}{k_c}  \right \|_2^2  \mbs J \right )^{-1}
&=&
 \sum_{p=0}^{+ \infty}  \left( \left \| \frac{\mbs k}{k_c}  \right \|_2^2  \mbs J \right )^p
 ,\\
&=&
 \sum_{p=0}^{+ \infty}  (-1)^p \left \| \frac{\mbs k}{k_c}  \right \|^{4p}_2  
 \id 
 +
 \sum_{p=0}^{+ \infty}  (-1)^p \left \| \frac{\mbs k}{k_c}  \right \|^{4p+2}_2  
   \mbs J
   ,\\
&=&
 \sum_{p=0}^{+ \infty}   \left( - \left \| \frac{\mbs k}{k_c}  \right \|^{4}_2 \right)^p 
 \left( \id + \left \| \frac{\mbs k}{k_c}  \right \|_2^2 \mbs J \right) 
   ,\\
&=&
 \frac 1 {1 + \left \| \frac{\mbs k}{k_c}  \right \|_2^4}
 \left( \id + \left \| \frac{\mbs k}{k_c}  \right \|_2^2 \mbs J \right) .
\end{eqnarray}
This leads to the following solution for the modified geostrophic balance:
\begin{eqnarray} 
\label{solution geostrophic modif}
\hat{ \mbs u}
=
 \frac 1 {1 + \left \| \frac{\mbs k}{k_c}  \right \|_2^4}
  \left( - \frac 1 f \hat {\mbs \xi}^{\bot} \right)
  +
 \frac {\left \| \frac{\mbs k}{k_c}  \right \|_2^2} {1 + \left \| \frac{\mbs k}{k_c}  \right \|_2^4}
   \left( \frac 1 f \hat {\mbs \xi} \right).
\end{eqnarray}


\section{Non-dimensional Boussinesq equations}
\label{appendix Nondimensionalized Boussinesq equations}

To derive a non-dimensional version of the  Boussinesq equations under location uncertainty \citep{resseguier2016geo1}, each term of the evolution laws is scaled \citep{resseguier2016geo2}:
the horizontal coordinates $\tilde \xx_h= L \xx_h$, the vertical coordinate $\tilde z=h z$, the aspect ratio $D=h/L$ between the vertical and horizontal length scales. A characteristic time $\tilde t=Tt$ corresponds to the horizontal advection time $U/L$ with horizontal velocity $\tilde{\mbs  u} =U\mbs u$. A vertical velocity $\tilde w=(h/L)U w$ is deduced from the divergence-free condition. We further take a scaled buoyancy $\tilde b = B b$, pressure $\tilde \phi'=\Phi \phi'$ (with the density scaled pressures $\phi'=p' / \rho_b$ and $\dif_t \phi_{\sigma}=\dif_t  p_{\sigma}/\rho_b$), and the earth rotation $\mbs f^*= f\mbs k$. For the uncertainty variables, we consider a horizontal uncertainty $\tilde {\mbs  a}_H =  A_u\;\mbs a_H$ corresponding to the horizontal $2\times 2$ variance tensor; a vertical uncertainty vector $\tilde{a}_{z z}= A_w { a}_{z z}$ and a horizontal-vertical uncertainty vector $\tilde{\mbs a}_{H z}= \sqrt{A_u A_w} {\mbs a}_{H z}$ related to the variance between the vertical and horizontal velocity components. The resulting non-dimensional Boussinesq system under location uncertainty becomes:

\fcolorbox{black}{lightgray}{
\begin{minipage}{0.95\textwidth}
\begin{center}
\bf  Nondimensional  Boussinesq equations under location uncertainty
\end{center}
\begin{subequations}
\label{sto-Boussinesq-buo}
\begin{align}
&\!\!\text{\em Momentum equations} \nonumber\\
\label{sto-Boussinesq-buo:Momentum equations}
&\;\;\;\;
\dif_t\mbs u 
+ \left(\w \bcdot \nab \right) \mbs u \dif t
+  \frac{1}{\Upsilon^{1/2} }\left( \bsigma_H \dBt  \bcdot \nab_H \right) \mbs u
+  \left( \frac \Ros {\Bu\Upsilon^{1/2}} \right)  (\bsigma \dBt)_z   \partial_z \mbs u
\nonumber \\ & \hspace{0.8cm}
-  \frac{1}{2 \Upsilon} \sum_{i,j\in H}  \partial_{ij}^2\biggl (a_{ij} \mbs u\biggr)  \dif t
+  O \left( \frac \Ros {\Upsilon \Bu} \right) 
+\frac{1}{\Ros}\left( f_0 + \Ros \beta y \right)  \mbs k  \times 
\left( \mbs u \dif t +  \frac{1}{\Upsilon^{1/2} }\bsigma_H \dBt \right)
\nonumber \\ & \hspace{0.8cm}
=
- \Eu\;\nab_H \left( \phi' \dif t + \frac{1}{\Upsilon^{1/2} } \dif_t \phi_\sigma \right) 
,\\
&\;\;\;\;
\dif_t w
+ \left(\w \bcdot \nab \right) w \dif t
+  \frac{1}{\Upsilon^{1/2} }\left( \bsigma_H \dBt  \bcdot \nab_H \right) w
+  \left( \frac \Ros {\Bu\Upsilon^{1/2}} \right)  (\bsigma \dBt)_z   \partial_z w
\nonumber
 \\ & \hspace{0.8cm}
-  \frac{1}{2 \Upsilon} \sum_{i,j\in H}  \partial_{ij}^2\biggl (a_{ij} w \biggr)  \dif t
+  O \left( \frac \Ros {\Upsilon \Bu} \right) 
=
\frac{\Gamma}{ D^2} b\dif t
  - \frac \Eu {D^2}\;\partial_z\left( \phi' \dif t + \frac{1}{\Upsilon^{1/2} } \dif_t \phi_\sigma \right) , 
\label{w-compad1}\\
&\!\!\text{\em Buoyancy equation} \nonumber\\
&\;\;\;\;
\nonumber
{\dif_t b} 
+  \left( \w_\Upsilon^*\dif t +   \frac{1}{\Upsilon^{1/2} }(\bsigma d\B_{t})\right)  
\bcdot \nab b
-\frac{1}{2}\frac{1}{\Upsilon} \nab_H \bcdot \bigl ( \mbs a_H \nab b \bigr ) \dif t
+ O \left( \frac \Ros {\Upsilon \Bu} \right) 
\\
& 
\hspace{4.5cm} +  \frac{1}{(\Fr)^2}\frac{1}{\Gamma} 
\left (
w^*_{\Upsilon/2} \dif t 
+ \left( \frac \Ros \Bu \right) \frac{1}{\Upsilon^{1/2}} (\bsigma \dif \B_t)_z
\right ) =0,\\
&\!\!\text{\em Effective drift} \nonumber\\
&\;\;\;\;
\w_{\Upsilon}^*
= 
\bigl(\mbs u^*_\Upsilon, w^*_\Upsilon \bigr)\transp ,
\nonumber\\
&\;\;\;\;\;\;\;\;
=
\left( 
\left( 
\w -\frac{1}{2\Upsilon} \nab\bcdot \mbs a_H
\right ),
\left( 
w 
- \left( \frac \Ros {2\Upsilon \Bu} \right)   \nab_H \bcdot {\mbs a}_{H z}
+ O \left( \frac \Ros {\Upsilon \Bu} \right)^2 
\right )
\right ) \transp,
\\
&\!\!\text{\em Incompressibility} \nonumber\\
&\;\;\;\;  \nab \bcdot \w =0,\\
&\;\;\;\; 
{\nab} {\bcdot}\bigl(\bsigma \dif \B_t\bigr)  = 0 ,\\
&\;\;\;\;  
 \nab_H \bcdot \left( \nab_H \bcdot \mbs a_H \right) \transp
 + 2 \frac \Ros \Bu \nab_H \bcdot \partial_z \mbs a_{Hz}
 + O \left( \left( \frac \Ros \Bu \right)^2 \right)
 =0.
\end{align}
\end{subequations}
\end{minipage}
} 
 \;\\\;\\

Here, the time-correlated components and the time-uncorrelated components in the momentum equations have not been separated.
The terms in $O\left( \frac \Ros \Bu \right)$ and $O\left( \frac \Ros \Bu \right)^2$ are related to the time-uncorrelated vertical velocity. These terms are too small to appear in the final QG model ($\Bu = O\left( 1\right)$ in QG approximation) and not explicitly shown. We only make appear the big $O$ approximations. Traditional non-dimensional numbers are introduced : the Rossby number $R_o= U/(f_0 L)$ with $f_0$ the average Coriolis frequency;
 the Froude number ($\Fr= U/(Nh)$), ratio between the advective time to the buoyancy time; $\Eu$, the Euler number,  ratio between the pressure force and the inertial forces, $\Gamma=Bh/U^2=D^2 B T/W$ the ratio between the mean potential energy to the mean kinetic energy. To scale the buoyancy equation, the ratio between the buoyancy advection and the stratification term has also been introduced:
 \begin{equation}
 \frac{B/T}{N^2W}
 =
 \frac{B}{N^2h}
 =
 \frac{U^2}{N^2h^2}
 \frac{Bh}{U^2}
 =
 Fr^2 \Gamma. 
 \end{equation}
 
 Besides those traditional dimensionless numbers, this system introduces $\Upsilon$, relating the large-scale kinetic energy to the energy dissipated by the unresolved component:
\begin{equation}
\Upsilon
= \frac{UL}{A_u }
= \frac{U^2}{A_u / T}.
\end{equation}


\section{QG model under strong uncertainty}
\label{appendix QG model under strong uncertainty}

For the case $\Upsilon$ close to the Rossby number, the diffusion term is not  negligible anymore and the geostrophic balance is modified. As the terms of the geostrophic balance remain large ($\Ros \leqslant \Upsilon \ll 1$), the scaling of the pressure can still be done with the Coriolis force. This leads to an Euler number scaling as
\begin{eqnarray}
\Eu \sim \frac 1 \Ros.
\end{eqnarray}
Keeping a small aspect ratio $D^2\ll1$, we get
\begin{eqnarray}
\frac {\Eu}{D^2} 
\sim  \frac 1{\Ros D^2} 
\gg \frac 1{\Ros}
\geqslant \frac 1{\Upsilon}.
\end{eqnarray}
As the Rossby number and the ratio $\Upsilon$ are both small in the vertical momentum equation, the inertial terms are dominated by the diffusion term which is itself negligible in front of the pressure term. The hydrostatic balance is hence conserved. The buoyancy scaling still correspond to the thermal winds relation:
\begin{eqnarray}
\Gamma \sim \Eu \sim \frac 1 \Ros.
\end{eqnarray}
Considering the scaling $\frac { (\bsigma \dif \B_t)_{z} }{\| (\bsigma \dif \B_t)_{H}\| } \sim D \frac \Ros \Bu$ for the vertical small-scale velocity,  the non-dimensional evolution equations are now given by:

\begin{align}
&\hspace*{-0.5cm}\text{\em Momentum equations} \nonumber\\
&
\label{scaled Boussinesq QG horiz momentum}
\Ros \left( 
\dif_t\mbs u 
+ \left(\mbs u \bcdot \nab \right) \mbs u \dif t
+  \frac{1}{\Upsilon^{1/2} }\left( \bsigma_H \dBt  \bcdot \nab \right) \mbs u
+  O \left( \frac \Ros {\Upsilon \Bu} \right) 
\right)
-  \frac{\Ros}{2 \Upsilon} \sum_{i,j\in H}  \partial_{ij}^2\bigl (a_{ij} \mbs u\bigr)  \dif t
\nonumber \\ & \hspace{0.8cm}
+\left( f_0 + \Ros \beta y \right)  \mbs k  \times 
\left( \mbs u \dif t +  \frac{1}{\Upsilon^{1/2} }\bsigma_H \dBt \right)
=
- \;\nab_H \left( \phi' \dif t + \frac{1}{\Upsilon^{1/2} } \dif_t \phi_\sigma \right) 
  ,\\
&
 b \; \dif t 
+  O \left( \frac{ \Ros D^2}{\Upsilon^{1/2}} \right) 
 = \;\partial_z\left( \phi' \dif t + \frac{1}{\Upsilon^{1/2} } \dif_t \phi_\sigma \right) , 
\label{scaled Boussinesq QG vert momentum}
\\
&\hspace*{-0.5cm}\text{\em Buoyancy equation} \nonumber\\
&
\label{scaled Boussinesq QG buoy}
\frac{ \Ros}{\Bu} \left( 
{\dif_t b} 
+ \nab b \bcdot \; \left( \mbs u \dif t +   \frac{1}{\Upsilon^{1/2} }(\bsigma d\B_{t})_H \right)  
+ \partial_z b \ w \dif t
\right)
-  \frac{\Ros}{2 \Upsilon} \sum_{i,j\in H}  \partial_{ij}^2\left(a_{ij} b \right)  \dif t
\nonumber \\ &
 \hspace{1.5cm}
+   w\dif t 
- \frac 1{ \Upsilon}  \frac \Ros \Bu \left ( \nab \bcdot a_{Hz}\right) \transp \dif t 
+  \frac \Ros \Bu \frac{1}{\Upsilon^{1/2}} (\bsigma \dif \B_t)_z
+ O \left( \frac {\Ros^2} {\Upsilon\Bu^2} \right) 
=0,
\\
&\hspace*{-0.5cm}\text{\em Incompressibility} \nonumber\\
&
 \nab \bcdot \mbs u+ \partial_z w =0,
 \label{QG deriv incomp1}\\
& 
 \nab {\bcdot}\bigl(\bsigma \dif \B_t\bigr)_H  
 + \frac \Ros \Bu \partial_z\bigl(\bsigma \dif \B_t\bigr)_z
 =0,
 \label{QG deriv incomp2}\\
& 
  \nab \bcdot \left( \nab \bcdot \mbs a_H \right) \transp
 + 2 \frac \Ros \Bu \nab \bcdot \partial_z \mbs a_{Hz}
 + O \left( \left( \frac \Ros \Bu \right)^2 \right)
 =0.
 \label{QG deriv incomp3}
\end{align}

The operators Del, $\nab$, and Laplacian, $\Delta$ represent $2$D operators. If $\Ros \sim \Upsilon$, the system is not anymore approximately in geostrophic balance. The large-scale velocity becomes divergent and decoupling the system is more involved. For sake of simplicity, we thus focus on the case of homogeneous and horizontally isotropic turbulence. As a consequence, the variance tensor $a$ is constant in space and diagonal:
\begin{equation}
a =
\begin{pmatrix}
a_h & 0 & 0 \\
0 & a_h & 0 \\
0 & 0 & a_z
\end{pmatrix}.
\end{equation}
The time-correlated components of the horizontal momentum at the $0$-th order can be written as:
\begin{eqnarray}
- \frac {a_H} 2 \Delta \mbs u_0 + f_0 \mbs k \times \mbs u_0 = - \nab \phi'_0,
\end{eqnarray}
Then, equation (\ref{solution geostrophic modif}) of Appendix \ref{appendix Modified geostrophic balance} expresses the result in Fourier space. In the physical space, the solution reads:
\begin{eqnarray} 
 \mbs u_0
=
 \nab^{\bot}
 \underbrace{
  \left( 1 +  \frac{\Delta^2}{k_c^4}  \right)^{-1}
\frac{\phi'_0 }{f_0}
}_{= \psi_0 }
  +
 \nab
 \underbrace{
 \left( 1 +  \frac{\Delta^2}{k_c^4}  \right)^{-1}
  \frac{\Delta}{k_c^2}
\frac{\phi'_0 }{f_0}
}_{= \tilde \psi_0}
\text{ with }
k_c = \sqrt{\frac{2f_0}{a_H}}
\end{eqnarray}
which is the Helmholtz decomposition of the horizontal velocity $u_0$ into its rotational and divergent component with a stream function $\psi_0$ and a velocity potential $\tilde \psi_0$. Differentiating the buoyancy equation at the order $0$ along z, we obtain
\begin{eqnarray}
\frac{a_H}{2} \Delta \partial_z \left( \frac{ b_0}{\Bu} \right)
=  \partial_z w_0 
= - \nab \bcdot \mbs u_0
=- \Delta \tilde \psi_0 
= - \frac {\Delta^2} {k_c^4}  \psi_0 .
\end{eqnarray}
The time-correlated part of the $0$-th order hydrostatic equation relates the buoyancy to the pressure $\phi_0'$:
\begin{eqnarray}
\frac{a_H}{2} \Delta \partial_z \left( \frac{ b_0}{\Bu} \right)
=
\frac{a_H}{2} \Delta \partial^2_z \phi_0'
=
\frac{a_H}{2} f_0 \Delta \partial^2_z \left(  1 +\frac {\Delta^2} {k_c^4}  \right ) \psi_0.
\end{eqnarray}
Gathering these two equations leads to:
\begin{eqnarray}
\left( \Delta 
+  
\left(  1 +\frac {\Delta^2} {k_c^4}  \right )
\partial_z
\left(
\left ( \frac{f_0}{N} \right)^2 
\partial_z
\right)
\right)
\psi
=  0.
\end{eqnarray}
Using the horizontal Fourier transform, it writes:
\begin{eqnarray}
\left( - \| \mbs k \|^2_2
+  
\left(  1 +  \left \| \frac {\mbs k} {k_c} \right \|^4_2  \right ) 
\partial_z
\left(
\left ( \frac{f_0}{N} \right)^2 
\partial_z
\right)
\right)
\hat \psi
=  0.
\end{eqnarray}
Under an uniform stratification, with a fixed value at a specific depth ($z=\eta$), and a vanishing condition in the deep ocean ($z \to - \infty$), a solution is:
\begin{equation}
\hat \psi (\mbs k,z)
=
\hat \psi (\mbs k,\eta) \ 
exp \left(
\frac{ N \| \mbs k \|_2}
{f_0
\sqrt{ 1 +  \left \| \frac {\mbs k} {k_c} \right \|^4_2 }
}
\ 
(z-\eta)
\right).
\end{equation}
Accordingly, the buoyancy is:
\begin{equation}
\hat b
=  \partial_z \hat \phi' 
= f_0 \left(
1 + \left \| \frac {\mbs k} {k_c} \right \|^4_2
\right) 
\partial_z \hat \psi 
=
N \| \mbs k \|_2
\sqrt{ 1 +  \left \| \frac {\mbs k} {k_c} \right \|^4_2 }
\ 
\hat \psi.
\end{equation}


\end{document}